\documentclass[a4paper,12pt]{article}

\usepackage[dvips]{graphicx}
\usepackage{bm}

\newcommand{\be}{\begin{equation}}
\newcommand{\ee}{\end{equation}}

\def\beq{\begin{equation}}
\def\eeq{\end{equation}}

\def\al{\alpha}

\def\Ga{\Gamma}

\def\de{\delta}
\def\De{\Delta}

\def\si{\sigma}
\def\Si{\Sigma}
\def\te{\theta}

\def\lam{\lambda}
\def\Om{\Omega}
\def\om{\omega}

\def\sq{\sqrt}

\def\l{\left (}
\def\r{\right )}

\def\fr{\frac}
\def\la{\label}
\def\hs{\hspace}
\def\vs{\vspace}

\def\ran{\rangle}
\def\lan{\langle}
\def\ov{\overline}

\def\tm{\times}

\begin{document}

\begin{flushright}
BA-06-19\\
OSU-HEP-06-13\\
December 22, 2006 \\
\end{flushright}

\vs{0.5cm}

\begin{center}
{\Large\bf

Missing Partner Mechanism in $SO(10)$ Grand Unification
}
\end{center}

\vspace{0.5cm}
\begin{center}
{\large
{}~K.S. Babu$^{a}$,
{}~I. Gogoladze$^{b}$,
{}~Z. Tavartkiladze$^{a}$
}
\vspace{0.5cm}

$^a${\em Department of Physics, Oklahoma State University, Stillwater, OK 74078, USA }

$^b${\em Bartol Research Institute, Department of Physics and Astronomy

University of Delaware, Newark, DE 19716, USA }

\end{center}

\begin{abstract}


We present a new possibility for achieving doublet--triplet splitting naturally in supersymmetric $SO(10)$ grand
unified theories. It is based on a missing partner mechanism which is realized with
the $126+\ov{126}$ Higgs superfields.  These Higgs fields, which are also needed for generating
Majorana right--handed neutrino masses, contain a pair of color triplets in excess of weak doublets.
This feature enables us to remove the color triplets from the low energy spectrum without fine--tuning.
We give all the needed ingredients for a successful implementation of the missing partner mechanism in $SO(10)$ and present explicit models wherein the Higgs doublet mass is protected against
possible non--renormalizable corrections to all orders.  We also show how realistic fermion masses can be
generated in this context.

\end{abstract}

\vs{0.7cm}


\newpage

\section{Introduction}

Unification of the different forces of Nature is a long sought--after dream of physicists.
Embedding the strong, weak, and electromagnetic forces into a single force represented by a Grand
Unified symmetry \cite{gut} provides a major step in this endeavor.  In such Grand Unified Theories
(GUT), not only are the Standard Model (SM) gauge interaction unified at high energies, but the matter
fields -- quark
and leptons -- are also unified, fitting into common multiplets of the GUT symmetry group.  This unification
of matter fields provides a simple understanding of the quantum numbers of quarks and leptons.
It is remarkable that with the inclusion of low energy supersymmetry, the SM gauge couplings, when
extrapolated from their
precisely known values at low energies, all meet at a common point at a scale of $M_X \simeq 2 \times 10^{16}$ GeV.
TeV scale supersymmetry (SUSY) is perhaps the most widely anticipated new physics for the LHC era,
since it provides a resolution to the gauge hierarchy problem.  It is no wonder that supersymmetric grand
unified theories have had a major influence in our thinking of physics beyond the Standard Model over
the last twenty five years.

Grand unification also predicts another, perhaps unwanted, unification: the Higgs doublet fields,
which must survive to low energies for spontaneous breaking of the electroweak gauge symmetry, are unified
with color triplet partners.  These color triplet Higgs fields must have masses of order the GUT scale, or else
they will upset the successful unification of gauge couplings, and will lead to proton decay with an
unacceptably large rate.  A major challenge for GUT model building is then to come up with a mechanism which
makes the electronweak Higgs doublets light while their color triplet partners remain heavy.  This is
perhaps the thorniest problem facing grand unified model building, and goes by the name doublet--triplet (DT)
splitting problem.

In SUSY $SU(5)$, which is the prototype for SUSY GUTs \cite{su5}, the Higgs doublets $h_{u,d}$
of MSSM are contained in $5+\overline{5}$ representations.  Under $SU(3)_c \times SU(2)_L \times U(1)_Y$, the $5$ breaks
up as $5 = h_u(1,2,1/2) + T(3, 1, -1/3)$, while $\overline{5} = h_d(1,2,-1/2)+\overline{T}(\overline{3}, 1, 1/3)$.
The superpotential for the model contains two terms involving the $5+\overline{5}$:
$W \supset M_5 5\cdot \overline{5}+ \lambda ~5\cdot \overline{5}\cdot 24$
where the $24$ is the Higgs field that breaks $SU(5)$ down to the SM gauge
symmetry with a vacuum expectation value $\left \langle 24 \right \rangle = V \times {\rm diag}.(1,1,1,-3/2,-3/2)$.
The doublet and triplet masses from this superpotential are then $m_D =M_5 -(3/2) \lambda V$ and $m_T = M_5 + \lambda V$
respectively.  If both $M_5$ and $V$ are of order the GUT scale, the color triplet will have
naturally mass of order the GUT scale.  In order to make the doublet mass at the weak scale a sever fine--tuning
between two large terms is done.  This  fine--tuning, which is present in other GUT groups as well,
is an unattractive aspect of SUSY GUTs.

Several mechanisms have been suggested in the literature for solving of the doublet--triplet splitting
problem without fine--tuning. In $SU(5)$, the missing partner mechanism \cite{mpSU5} (briefly reviewed
in the next section) can be employed to avoid the fine--tuning.  Here one makes use of  $50+\overline{50}$
Higgs fields which have the feature that they  contain color triplets, but no weak doublets.
One pairs up the colored higgses from the $5+\overline{5}$ with those from the $50+\overline{50}$.  Such
a pairing will leave the doublets naturally light. The stability  of such a solution against higher order operators requires some additional effort \cite{Berezhiani:1996nu}.

In $SU(6)$ grand unified theories, the pseudo--Goldstone mechanism \cite{Berezhiani:1989bd, pgb-anU1}
can solve the DT splitting problem rather elegantly.  Here the Higgs doublets are identified
as pseudo--Goldstone bosons of a larger global symmetry.  The gauge symmetry should be augmented
by additional symmetries for this realization. The anomalous ${\cal U}(1)$ symmetry of string origin
is very efficient for this purpose \cite{pgb-anU1}.

In this paper we are concerned with grand unified theories based on SUSY $SO(10)$, which are
 particularly attractive \cite{so10}.
The spinor representation of $SO(10)$ unifies
all matter fermions of a given family in a single multiplet -- a feat not achieved in $SU(5)$
or $SU(6)$ GUT.  The spinor of $SO(10)$  contains also the right handed neutrino ($\nu_R$), which can generate light neutrino masses via see-saw mechanism \cite{seesaw}. The $\nu_R$ can also naturally account for the
baryon asymmetry of the Universe via leptogenesis \cite{Fukugita:1986hr}.
Another nice property of $SO(10)$--based GUT is that, they can automatically lead to matter parity (or $R$--parity) \cite{rp} which is usually assumed as an ad hoc symmetry in the MSSM.  Such a symmetry is needed in order to avoid
rapid proton decay, it also provides a natural cold dark matter candidate.  In SUSY $SO(10)$, matter parity
can be automatic since it contains $B-L$ as a subgroup.

The most widely discussed approach to the DT splitting problem in $SO(10)$ is the Dimopoulous--Wilczek
mechanism, or the missing VEV mechanism \cite{Dimopoulos:1981xm}.  Here one employs an adjoint
45--plet of Higgs field with its vacuum expectation value pointing in the $B-L$ conserving direction:
$\left\langle 45 \right \rangle = V \times {\rm diag}(1,1,1,0,0) \otimes i\tau_2$.  The MSSM Higgs doublets
are contained in $10$ of $SO(10)$ ($10 = 5+\overline{5}$ under $SU(5)$).
If the superpotential contains the terms $W \supset M_{10} 10'10'+
\lambda ~10\hs{-0.4mm}\cdot \hs{-0.5mm}10'\hs{-0.4mm}\cdot \hs{-0.4mm} 45$, because of the VEV structure of $45$, the color triplets from $10, 10'$ will acquire
GUT scale masses, while a  doublet pair from $10$ will remain massless.  Note that this is done without
fine--tuning, and is facilitated by the fact that the adjoint of $SO(10)$ is not a traceless matrix, unlike
the adjoint of $SU(N)$ groups.  A variety of realistic models based on this mechanism have been constructed
in the literature \cite{babubarr2}.  Additional ingredients are usually needed to guarantee the stability of the VEV structure of $45$ \cite{raby,so10-DT}.  Realistic models for fermion masses including neutrino
oscillations have been constructed based on this mechanism \cite{pati}.

Although $SO(10)$--based model building has attracted considerable attention, to date, the missing partner
mechanism that works in SUSY $SU(5)$ has not been successfully implemented in $SO(10)$.  The purpose of this paper is
provide such a realization. We present examples of models  with all order stability of the DT hierarchy which
also have realistic phenomenology.

In the next sections, first we point out the possibility and some properties of a missing doublet $SO(10)$ scenario and draw the ingredients needed for realistic model building. Then we present explicit models which provide all order stability of the proposed DT splitting mechanism.  Then we show how realistic fermion masses can be generated in this
context.  Finally we comment on the perturbativity of the gauge coupling above the GUT scale.

\section{Missing Partner Mechanism in $SO(10)$}

First let us recall how the missing partner mechanism, or missing doublet (MD) mechanism,
works within SUSY $SU(5)$ GUT. Then, the steps needed for building a missing doublet $SO(10)$
(MDSO10) GUT will be easier to follow. In SUSY $SU(5)$, the pair of the MSSM Higgs doublets $h_u$ and $h_d$ are embedded in the supermultiplets $h(5)$ and $\bar h(\bar 5)$ respectively. The composition of these
states are $h=h_u+T_h$, $\bar h=h_d+\bar T_h$, where $T, \bar T$ are color triplets.  The $50+\overline{50}$ representations
of $SU(5)$  (which we denote as $\psi + \bar \psi$) have the curious feature that they
contain states with the same quantum numbers as $T_h$ and $\bar T_h$, but not the $h_u, ~h_d$
states  \cite{Slansky:1981yr}. Thus arranging suitable couplings between $h, \bar h$ and $\psi , \ov{\psi }$ one can decouple the triplets $T_h, \bar T_h$ from $5 + \overline{5}$ through the mixing with $T_{\psi }, \bar T_{\psi }$ \cite{mpSU5}.
For this to be achieved, a scalar $\phi (75)$-plet must be introduced with non--zero
$SU(5)\to SU(3)_c \times SU(2)_L \times U(1)_Y$ breaking VEV.
The relevant  superpotential couplings are
\beq
W \supset \phi h\ov{\psi }+\phi \bar h\psi +M_{\psi }\ov{\psi }\psi ~.
\la{MDSU5}
\eeq
After substituting the VEV $\lan \phi \ran \sim V_{\phi }$, the color triplet mass matrix will be

\beq
\begin{array}{cc}
 & {\begin{array}{cc}
\hs{-0.2cm} \bar T_h\hspace{0.2cm} & \hspace{0.1cm} \bar T_{\psi }

\end{array}}\\ \vspace{1mm}
\begin{array}{c}
 \vs{0.1cm}T_h\\ T_{\psi }
 \end{array}\!\!\!\!\!\!\!\hs{0cm} &{\left(\begin{array}{cc}

\hs{0.3mm} 0&\hs{0.1cm} V_{\phi }
\\
\vs{-0.4cm}
\\
 \hs{0.3mm}V_{\phi }& \hs{0.1cm} M_{\psi }
\\
 &
 \vs{-0.4cm}

\end{array}\right)},
\end{array}
\la{M-T-SU5}
\eeq
and therefore with $M_{\psi }\sim V_{\phi }\sim M_{\rm GUT}$ one expects all triplet masses to be near the scale  $M_{\rm GUT}$. At this level the doublets from $5+\overline{5}$ are massless since there are no doublets in
the $50+\overline{50}$.   Crucial for this mechanism is the omission of the mass term $M_h\bar hh$. Of course, this must be justified and additional symmetries can be employed for this purpose \cite{Berezhiani:1996nu}.

Now we turn to models based on $SO(10)$ gauge symmetry and try to
see how the missing partner mechanism can be realized. The lowest dimensional Higgs representations which has
a missing doublet in $SO(10)$ is the $126+ \overline{126}$. They contain $SU(5)$ $\ov{50}+\overline{50}$-plets. Indeed, in terms of $SU(5)\tm U(1)\equiv G_{51}$ (one of the maximal subgroups of $SO(10)$) we have \cite{Slansky:1981yr}
\beq
126=1_{-10}+\bar 5_{-2}+10_{-6}+\ov{15}_6+45_2+\ov{50}_{-2}~,
\la{dec126}
\eeq
where the subscripts stand for $U(1)$ charges. This representation, together with $\ov{126}$, will be used for building MDSO10 model. The MSSM Higgs doublets are embedded (at least partially) in the scalar $H(10)$-plet of $SO(10)$. This is the lowest representation which admits renormalizable Yukawa couplings -
$16\cdot 16~ H(10)$. So, for DT splitting we should arrange the coupling of $H$ with $126$-plets. To do this at the renormalizable level we need to introduce a scalar supermultiplet in the $210$
representation of $SO(10)$. Note that it is quite exciting that the multiplets $126+\ov{126}+210$, which we just mentioned, have been used extensively recently  for building
renormalizable $SO(10)$ GUT \cite{n-minSO101, n-minSO102} with some predictive power in the
fermion sector, including neutrino oscillations.

Here we will show the importance of these states in achieving the DT splitting. The couplings of the bi-linears $126~H(10)$
and $\ov{126}~H(10)$  with $210$-plet form  $SO(10)$ singlets and will be used below. However, we note that some caution is needed for building self--consistent model. From Eq. (\ref{dec126})
we see that the $126, \ov{126}$-plets contain, besides the $50$-plets, $5$ and $45$-plets. The latter states
contain color triplets as well as weak doublets. Thus there is danger that together
with color triplets of $H(10)$ the doublets also gain large masses. However, if we introduce a set of Higgs superfields containing in total three pair of weak doublets, it is clear by a simple
counting of degrees of freedom that, only two pair of doublets will get masses by mixing with $5$ and $45$-plets from the $126, \ov{126}$. Now, one must decide which additional states are
most convenient for this purpose together with one $H(10)$ supermultiplet.
It turns out that the state $\Si (120)$ can do a very useful job in this regard. The decomposition of
$120$ in terms of $G_{51}$ reads
\beq
120=5_2+\bar 5_{-2}+10_{-6}+\ov{10}_6+45_2+\ov{45}_{-2}~.
\la{dec120}
\eeq
The multiplets $H(10)$ and $\Si (120)$ together contain three pairs of doublets and three triplet-antitriplet pairs.
To be more clear, let us consider the multiplets $H(10), \Si (120), \De (126), \bar{\De }(\ov{126}), \Phi (210)$ and the superpotential
couplings
\beq
\Phi \De (H+\Si )+\Phi \bar{\De } (H+\Si )+M_{\De }\De \bar{\De }~.
\la{toy-W-DT}
\eeq
With $\lan \De \ran =\lan \bar{\De }\ran =0$  and the VEV of $\Phi $ in the most general direction
that preserves the SM gauge symmetry (see
next section for more details) the mass matrices for triplet and doublet states schematically are given by

\beq
\begin{array}{cc}
 & {\begin{array}{cc}
\hs{-0.6cm} {\bf \bar T}_{H, \Si }\hspace{0.4cm} & \hspace{0.3cm} {\bf \bar T}_{\De }

\end{array}}\\ \vspace{1mm}
{\bf M_T}=
\begin{array}{c}
 \vs{0.1cm}{\bf T}_{H, \Si }\\ {\bf T}_{\De }
 \end{array}\!\!\!\!\!\!\!\hs{0cm} &{\left(\begin{array}{cc}

\hs{0.3mm} {\bf 0}_{3\tm 3}&\hs{0.1cm} \lan {\bf \Phi }\ran_{3\tm 3}
\\
\vs{-0.4cm}
\\
 \hs{0.3mm}\lan {\bf \Phi }\ran_{3\tm 3} & \hs{0.1cm} \l {\bf M_{\De }}\r_{3\tm 3}
\\
 &
 \vs{-0.4cm}

\end{array}\right)},
\end{array}
\la{toy-MT}
\eeq

\beq
\begin{array}{cc}
 & {\begin{array}{cc}
\hs{-0.6cm} {\bf \bar D}_{H, \Si }\hspace{0.4cm} & \hspace{0.3cm} {\bf \bar D}_{\De }

\end{array}}\\ \vspace{1mm}
{\bf M_D}=
\begin{array}{c}
 \vs{0.1cm}{\bf D}_{H, \Si }\\ {\bf D}_{\De }
 \end{array}\!\!\!\!\!\!\!\hs{0cm} &{\left(\begin{array}{cc}

\hs{0.3mm} {\bf 0}_{3\tm 3}&\hs{0.1cm} \lan {\bf \Phi }\ran_{3\tm 2}
\\
\vs{-0.4cm}
\\
 \hs{0.3mm}\lan {\bf \Phi }\ran_{2\tm 3} & \hs{0.1cm} \l {\bf M_{\De }}\r_{2\tm 2}
\\
 &
 \vs{-0.4cm}

\end{array}\right)},
\end{array}
\la{toy-MD}
\eeq
where the dimensions of the block matrices have been denoted appropriately by subscripts. The dimension of the doublet mass matrix is by one unit less than the dimension of the triplet
mass matrix, because there is one missing doublet pair (in states $50_{\bar{\De }}+\ov{50}_{\De }$).
Considering now the matrix Eq. (\ref{toy-MT}), we can see that all triplets from $H, \Si $-plets gain masses through the mixings with the three triplet--antitriplet pairs from
$\De (126)+\bar{\De }(\ov{126})$-plets. However, according to Eq. (\ref{toy-MD}), two doublet pairs from  $\De (126)+\bar{\De }(\ov{126})$ generate masses for two doublet pairs from $H, \Si $ states. Therefore, the third pair of doublets coming from $H, \Si $ will remain massless.
(This is also obvious from the fact that ${\rm Det}({\bf M_D})=0$). The reason is simple: as already was mentioned, there is one {\it missing} pair of doublet in  $\De (126)+\bar{\De }(\ov{126})$.

This is a transparent demonstration how the missing partner mechanism can work in $SO(10)$. However, for realistic model building some more elaboration will be required. Namely, one should make
sure that the couplings $H^2, \Si^2, \Phi H\Si $ are absent, i.e. the zero of the $3\tm 3$ block in  Eq. (\ref{toy-MD})
must be guaranteed. Although the supersymmetric non-renormalization theorem  guarantees that once set to zero
these terms will not be generated perturbatively, we wish to explain the origin of their absence based on some symmetries. In addition, 
we wish to insure that certain higher order non-renormalizable operators which may be induced by unknown Planck scale effects are absent.  Also, in order the for the DT hierarchy to remain intact we need the VEV of either $\De $ or $\bar{\De }$ to be zero. In the next section we present explicit $SO(10)$ model(s) which address all these issues and show the consistency of the
mechanism.

Before closing this section, let us mention that besides the triplet and doublet states $\Si (120)$-plet contain
other vector like states. All of these extra states will acquire
masses through the mixings with $\De (126), \bar{\De }(\ov{126})$ multiplets. Note that
the quantum numbers of all the fragments of $\Si (120)$-plet match with those of the states
from $\De ,\bar{\De }$ [see Eq. (\ref{dec126}) and Eq. (\ref{dec120})]. Therefore, no state (besides the one massless doublet which partially also resides in $H$) from $\Si $ remains massless.

\section{Explicit Missing Doublet $SO(10)$ Models}

 From the discussions of the previous section we already got a clear idea of what field content we
 would  need in order to realize the missing doublet mechanism in $SO(10)$.
As it will turn out, it is much more convenient if the $126$-plets involved in the DT splitting have no VEVs (or at least one, out of $\De (126)$ and $\bar{\De }(\ov{126})$-plets,
has no VEV). Thus, the symmetry breaking sector should be discussed is some detail. For the rank breaking of $SO(10)$ we can use either a scalar $16+\overline{16}$-plets or another
$126 + \overline{126}$-plets. In either case we denote the rank breaking superfields by $C, \bar C$ and distinguish between two possible cases:
\beq
{\bf (a)}~:~~~~C=16~,~~~~\bar C=\ov{16}~,~~~~~~{\rm and}~~~~~~{\bf (b)}~:~~~~C=126~,~~~~\bar C=\ov{126}~.
\la{C-cases}
\eeq
The states $C, \bar C$ together with $\Phi (210)$-plet break the $SO(10)$ group down to $SU(3)_c\tm SU(2)_L\tm U(1)_Y$. This discussion concludes the selection of the GUT scalar
superfields.

We wish to build models which preserves DT splitting to all order, i.e. all couplings (including
non--renormalizable operators) allowed by symmetries must be taken into account. Thus, we will need to forbid some of the couplings and the easiest way to do so is to introduce an additional gauge ${\cal U}(1)$ symmetry. As it  turns out, this symmetry is anomalous. The anomalous ${\cal U}(1)$ symmetry of string origin has been applied in
GUT model building
\cite{pgb-anU1, Berezhiani:1996nu, so10-DT} and has been shown to be very efficient for stabilizing the DT splitting
to all orders. Here we apply this ${\cal U}(1)$ symmetry in our MDSO10 scenario.
The anomalous $U(1)$ factors can appear in effective field theories from string theory upon compactification
to four dimensions.  The apparent anomaly in this ${\cal U}(1)$ is canceled
through the Green-Schwarz mechanism \cite{Green:1984sg}. Due to the
anomaly, a Fayet-Iliopoulos term $-\xi \int d^4\te V_A$ is always generated \cite{Witten:1981nf} and the corresponding
$D_A$-term  has the form \cite{Dine:1987xk}
\beq
\fr{g_A^2}{8}D_A^2=\fr{g_A^2}{8}\l -\xi +\sum Q_i|\phi_i|^2\r^2~,~~~\xi =\fr{g_A^2M_P^2}{192\pi^2}{\rm Tr}Q~,
\la{FI-D-A}
\eeq
where $Q_i$ is the ${\cal U}(1)$ charge of $\phi_i$ superfield. For ${\cal U}(1)$ breaking we introduce an $SO(10)$ singlet scalar superfield
$X$ with ${\cal U}(1)$ charge $Q_X=2$. With $\xi >0$, in Eq. (\ref{FI-D-A}) the VEV of the scalar component of
$X$ is fixed as $\lan X\ran =\sq{\xi/2}$.

In Table \ref{t:1} we list all scalar superfields introduced, the matter $16_i$-plets ($i=1,2,3$) and the corresponding ${\cal U}(1)$ charges.
%
%
%
\begin{table} \caption{${\cal U}(1)$ charges  $Q$ of the superfields. In case {\bf (a)}: $C=16, \bar C=\ov{16}$. In case {\bf (b)}: $C=126, \bar C=\ov{126}$.
 }

\label{t:1} $$\begin{array}{|c||c|c|c|c|c|c|c|c|c|}

\hline
\vs{-0.2cm}
 &  &  &  & & & & & &\\

\vs{-0.3cm}

& ~X~& ~H(10)~&~ \Si (120)~& ~\De (126)~&~ \bar{\De }(\ov{126}) ~  & ~\Phi (210) ~ & ~C ~  &  ~\bar C ~ & ~16_i\\

&  &  &  & &  & & & &\\

\hline
\vs{0cm}
&  &  &  & &   & &{\bf (a)}:3/2 &{\bf (a)}: -3/2 &\\

\vs{-0.12cm}

~Q~& 2 & 1 &1&-1 &-1 &0 & & &-\fr{1}{2}\\

&  &  &  & & & & {\bf (b)}:~3~~&{\bf (b)}:~-3~&\\

\hline

\end{array}$$

\end{table}
%
%
The fermion sector will be discussed at the end of this section.
With this assignment we can write down the superpotential couplings. The part which is important for DT splitting is
\beq
W_{\rm DT}=\Phi \De \l H+\Si \r +\Phi \bar{\De }\l H+\Si \r+X\bar{\De }\De ~.
\la{W-DT}
\eeq
In order to carry the detailed analysis, we should first investigate the symmetry breaking and field VEV structure.
The superpotential couplings important for the symmetry breaking are
\beq
W(\Phi, C)=\fr{\lam }{3}\Phi^3+\fr{M_{\Phi }}{2}\Phi^2+\bar CC\l M_C+\si \Phi \r ~.
\la{W-Phi-C}
\eeq
Also, there are higher order superpotential couplings (potentially induced by unknown gravity
effects) with `$\De -C$ mixing':
\beq
W(\De ,C)=
\left\{ \begin{array}{lll}
~X^2\De \bar C\bar C/M_{\rm Pl}^2~, &~ {\rm for~case}~ {\bf (a)} \\
 ~X^2\De \bar C/M_{\rm Pl}~,  &~ {\rm for~case}~ {\bf (b)}
\end{array}
\right.~.
\la{W-De-C}
\eeq
Thus, the total symmetry breaking  superpotential is
\beq
W_{\rm SB}=W(\Phi , C)+W(\De , C)~.
\la{W-SB}
\eeq

In terms of $SU(5)$ group, $\Phi (210)$ decomposes as
\beq
\Phi (210)=1_{\Phi }+24_{\Phi }+75_{\Phi }+\cdots
\la{dec210}
\eeq
where the dots stand for states which have no $SU(3)_c \times SU(2)_L \times U(1)_Y$ singlet components. Thus, only the first three fragments of $\Phi $
given in Eq. (\ref{dec210}) are relevant in studying the VEV structure. For denoting their VEVs we will introduce the following
notations
\beq
\Phi_1=\lan 1_{\Phi }\ran ~,~~~~\Phi_{24}=\lan 24_{\Phi }\ran ~,~~~~\Phi_{75}=\lan 75_{\Phi }\ran ~.
\la{210-VEVs}
\eeq
The VEVs of $C, \bar C$ will wind towards the $SU(5)$ singlet direction and will be denoted as
\beq
C_1=\lan C\ran ~,~~~~~\bar C_1=\lan \bar C\ran ~.
\la{CCbar-VEVs}
\eeq
Similarly, the $SU(5)$ singlet fragments in $\De $ and  $\bar{\De }$ can have (induced) VEVs and will be denoted
as $\De_1$ and $\bar{\De }_1$ respectively. For completeness we will take these induced VEVs also into account.
From the $F$--flatness  conditions $F_X=F_{\De }=F_{\bar{\De }}=0$ we have the solution
\beq
\De_1=0~,~~~\bar{\De }_1=\lan X\ran \de ~,~~~~{\rm with}~~~
\de\sim \left\{ \begin{array}{lll}
~\bar C_1^2/M_{\rm Pl}^2~, &~ {\rm for~case}~ {\bf (a)} \\
 ~\bar C_1/M_{\rm Pl}~,  &~ {\rm for~case}~ {\bf (b)}
\end{array}
\right.~.
\la{Sol-De-VEVs}
\eeq
On the other hand, $D$--flatness conditions for the anomalous ${\cal U}(1)$ and the $U(1)$ of $SO(10)$ are:
$$
-\xi +2|X|^2+Q_C\l |C_1|^2-|\bar C_1|^2\r -|\De_1|^2-|\bar{\De }_1|^2=0
$$
\beq
q_{U}\l |C_1|^2-|\bar C_1|^2\r -10|\De_1|^2+10|\bar{\De }_1|^2=0~,
\la{D-fl}
\eeq
where the ${\cal U}(1)$ charge $Q_C$ is given in Table \ref{t:1}, while $q_U=-5$ and $-10$ for cases {\bf (a)} and {\bf (b)} respectively.
In addition, from the $F_C=F_{\bar C}=0$ conditions we fix
\beq
\Phi_1\sim -\fr{M_C}{\si }~,
\la{Phi1-VEV}
\eeq
while the condition $F_{\Phi }$ gives schematically (up to some irrelevant Clebsch factors)
$$
\lam \l \Phi_1^2+\Phi_{24}^2+\Phi_{75}^2\r +M_{\Phi }\Phi_1 +\si \bar C_1C_1=0~,
$$
$$
\lam \l \Phi_{24}^2+\Phi_{1}\Phi_{24}+\Phi_{24}\Phi_{75}+\Phi_{75}^2\r +M_{\Phi }\Phi_{24}=0~,
$$
\beq
\lam \l \Phi_{75}^2+\Phi_{24}^2+\Phi_{1}\Phi_{75}+\Phi_{24}\Phi_{75}\r +M_{\Phi }\Phi_{75}=0~.
\la{F-Phi-fl}
\eeq
One can easily verify that the conditions Eqs. (\ref{Sol-De-VEVs})-(\ref{F-Phi-fl}) fix non--zero $\lan X\ran , \Phi_{1, 24, 75}$ and
$C_1, \bar C_1$ VEVs. For simplicity we can assume that all this VEVs are $\sim M_{\rm GUT}$. We also have $\bar{\De }_1\neq 0$, $\De_1=0$, and all the $F$ and the $D$-terms vanish, ensuring unbroken supersymmetry.  It is
important that the operators $H^2, \Si^2, \Phi H\Si $ are forbidden by ${\cal U}(1)$ symmetry. ${\cal U}(1)$ invariance would require
that these operators should be multiplied by some field combinations carrying negative ${\cal U}(1)$ charge. We can readily check  that such operator will
involve $\De $ and since $\lan \De \ran =0$ they are not relevant. Therefore, quadratic couplings with respect to $H, \Si $ will not give rise to
the doublet masses to all orders. There will be additional operators which are linear with respect to $H$ and $\Si$:
\beq
{\cal O}_HH+{\cal O}_{\Si }\Si~.
\la{H-Si-lin}
\eeq
The $SO(10)\tm {\cal U}(1)$ symmetry determines their structure and for the two cases {\bf (a)} and {\bf (b)} we have
$$
{\bf (a)}~:~~~{\cal O}_H=\De \bar C^2C^2+X\bar C^2~,~~~{\cal O}_{\Si }=\De \bar C^2C^2+X\Phi \bar C^2~,
$$
\beq
{\bf (b)}~:~~~{\cal O}_H=\De \Phi \bar CC+X\Phi \bar C~,~~~{\cal O}_{\Si }=\De \Phi \bar CC+X\Phi \bar C~,
\la{form-QHSi}
\eeq
(cut off scale is ommited).
As we will see, these operators do not spoil the DT hierarchy. We will take them into account in order to demonstrate
that we are getting successful DT splitting.

Since the VEV configuration and all superpotential terms are already fixed, we are ready to discuss the issue of the DT splitting.
The relevant coupling matrix in terms of $SU(5)$ fragments is

\begin{equation}
\begin{array}{cccccc}
 & {\begin{array}{ccccccc}
\hs{0.4cm} 5_H\hspace{0.4cm} & \hspace{0.5cm}5_{\Si } \hspace{0.3cm}
& \hspace{0.5cm}45_{\Si }\hspace{0.3cm} &\hspace{0.6cm}{\bf f}_C\hspace{0.5cm} &
\hspace{0.5cm}{\bf F}_{\De }\hspace{0.4cm}  &\hs{0.5cm}5_{\Phi } \hspace{0.3cm} &~
\end{array}}\\ \vspace{1mm}
\begin{array}{c}
\bar 5_H\\ ~\\ \vs{0.1cm} \bar 5_{\Si } \\ ~ \\ \ov{45}_{\Si } \\ ~\\ {\bf \bar f}_C \\ ~\\ {\bf \bar F}_{\De }\\ ~\\ \bar 5_{\Phi }
 \end{array}\!\!\!\!\! \hs{-0.1cm}&{\left(\begin{array}{cccccc}

 &&&&&
\\
\hs{0.3cm}0  &\hs{1.1cm}0  &\hs{1cm}0& ~& ~&
\\
 &&&&&
 \\
\hs{0.3cm}0  &\hs{1cm}0  &\hs{1cm}0&
\hs{1cm} {\bf \Ga } & \hs{1cm}{\bf \Om }  & \hs{1cm}{\bf \om }~
\\
 &&&&&
\\
\hs{0.3cm}0  &\hs{1.1cm}0  &\hs{1cm}0&~  & ~ &
\\
 &&&&&
\\
~ & \hs{1.1cm}0 &  ~& \hs{1cm}{\bf M_f}& \hs{1cm}0 &\hs{1cm}{\bf v}
\\
 &&&&&
\\
 ~ & \hs{1.1cm}{\bf \bar{\Om }} &  ~& \hs{1cm}{\bf q}& \hs{1cm}{\bf M_F} &\hs{1cm}0
 \\
 &&&&&
\\
  ~ & \hs{1.1cm}0 &  ~& \hs{1cm}{\bf \bar v}& \hs{1cm}0 &\hs{1cm}M_{\Phi }
 \\
 &&&&&
\end{array}\hs{-0.1cm}\right)}~,
\end{array}  \!\!
\label{M-DT}
\end{equation}
where each subscript indicates where the appropriate superfield fragment is coming from. For the `vector' states the following
notations have been used:
\beq
{\bf F}_{\De }=\l 5_{\bar{\De }},~45_{\De },~50_{\bar{\De }}\r ~,~~~~
{\bf \bar F}_{\De }=\hs{-0.3cm}
\begin{array}{cc}
\vspace{1mm}

\begin{array}{c}
\vs{0.1cm}\\
 \end{array}\!\!\!\!\!\hs{-0.2cm} &{\left(\hs{-0.15cm}
 \begin{array}{ccc}
\bar 5_{\De }
\\
\ov{45}_{\bar{\De }}
\\
\ov{50}_{\De }

\end{array}\hs{-0.15cm}\right)}\vs{-0.2cm}~,
\end{array}
\label{Fs}
\end{equation}
and for the two cases we have:
$$
{\bf (a)}:~~~{\rm For}~~~C=16~,~\bar C=\ov{16}~,~~~~~{\bf f}_C=5_{\bar C}~, ~~~~{\bf \bar f}_C=\bar 5_C
$$
\beq
{\bf (b)}:~~~{\rm For}~~~C=126~,~\bar C=\ov{126}~,~~~~~{\bf f}_C=\l 5_{\bar C},~45_C,~50_{\bar C}\r ~,~~~~
{\bf \bar f}_C=\hs{-0.3cm}
\begin{array}{cc}
\vspace{1mm}

\begin{array}{c}
\vs{0.1cm}\\
 \end{array}\!\!\!\!\!\hs{-0.2cm} &{\left(\hs{-0.15cm}
 \begin{array}{ccc}
\bar 5_C
\\
\ov{45}_{\bar C}
\\
\ov{50}_C

\end{array}\hs{-0.15cm}\right)}\vs{-0.2cm}~.
\end{array}
\label{fs}
\end{equation}
The blocks appearing in (\ref{M-DT}) are given by
\beq
\begin{array}{ccc}
 & {\begin{array}{ccc}
 & &
\end{array}}\\ \vspace{1mm}
{\bf \Om }\propto
 &{\left(\begin{array}{ccc}

\hs{0.3mm} \Phi_1+\Phi_{24}&\hs{0.1cm}\Phi_{24}+\Phi_{75}  &\hs{0.1cm} \Phi_{75}
\\
\vs{-0.4cm}
\\
 \hs{0.3mm}\Phi_1+\Phi_{24} & \hs{0.1cm} \Phi_{24}+\Phi_{75}& \hs{0.1cm} \Phi_{75}
 \\
 \vs{-0.4cm}
 \\
 \hs{0.3mm} \Phi_{24}+\Phi_{75}&\hs{0.1cm}\Phi_1+\Phi_{24}+\Phi_{75}&\hs{0.1cm} \Phi_{24}+\Phi_{75}

\end{array}\right)},~~~
{\bf \om }\propto \hs{-0.3cm}
\begin{array}{cc}
\vspace{1mm}

\begin{array}{c}
\vs{0.1cm}\\
 \end{array}\!\!\!\!\!\hs{-0.2cm} &{\left(\hs{-0.15cm}
 \begin{array}{ccc}
\bar{\De }_1
\\
 \bar{\De }_1
\\
(\Phi_{24}+\Phi_{75})\bar{\De }_1/M_{\rm Pl}

\end{array}\hs{-0.15cm}\right)}\vs{-0.2cm}~,
\end{array}
\end{array}
\la{Om-om}
\eeq
and ${\bf \bar{\Om }}$ has the structure of ${\bf \Om }^T$, while ${\bf M_F}\propto \lan X\ran~ {\rm diag }\l 1~,1~,1\r $.
Forms of ${\bf \Ga }, {\bf M_f}$, ${\bf q}$, ${\bf v}$ and ${\bf \bar v}$ depend on  the case we are dealing with [either {\bf (a)} or  {\bf (b)}].
For example, for case {\bf (a)}, i.e. when the rank reduction occurs
by $C(16), \bar C(\ov{16})$-plets, we have
\beq
{\bf \Ga }\propto \hs{-0.3cm}
\begin{array}{cc}
\vspace{1mm}

\begin{array}{c}
\vs{0.1cm}\\
 \end{array}\!\!\!\!\!\hs{-0.2cm} &{\left(\hs{-0.15cm}
 \begin{array}{ccc}
\bar C_1
\\
 \bar C_1
\\
(\Phi_{24}+\Phi_{75})\bar C_1/M_{\rm Pl}

\end{array}\hs{-0.15cm}\right)}\fr{X}{M_{\rm Pl}}\vs{-0.2cm}~,
\end{array}~,~~~~~
{\bf q}\propto \hs{-0.3cm}
\begin{array}{cc}
\vspace{1mm}

\begin{array}{c}
\vs{0.1cm}\\
 \end{array}\!\!\!\!\!\hs{-0.2cm} &{\left(\hs{-0.15cm}
 \begin{array}{ccc}
\bar C_1
\\
0
\\
\Phi_{75}\bar C_1/M_{\rm Pl}

\end{array}\hs{-0.15cm}\right)}\l \fr{X}{M_{\rm Pl}}\r^2\vs{-0.2cm}~,
\end{array}
\la{Ga-q}
\eeq
\beq
{\bf M_f}=M_C~,~~~{\bf v}\sim \bar C_1~,~~~{\bf \bar v}\sim C_1~,
\la{Mf-v-bv}
\eeq
where $M_C$ is the mass of $5$-plets from $C, \bar C$ arising from symmetry breaking superpotential. These block entries have different dimensions for case {\bf (b)}.
However, it is remarkable that the result does not depend on the structure of these entries. This becomes obvious from the whole form of the matrix  Eq. (\ref{M-DT}).
The integration of the states ${\bf f}_C, {\bf \bar f}_C$ and $5_{\Phi }, \bar 5_{\Phi }$ does not give any contribution to the upper left $3\tm 3$ zero block matrix of Eq. (\ref{M-DT}).
An important role for this is played by the off--diagonal zero block matrices which are protected by ${\cal U}(1)$ symmetry.
Upon integration of ${\bf f}_C, {\bf \bar f}_C, 5_{\Phi }, \bar 5_{\Phi }$ states the matrix  Eq. (\ref{M-DT}) reduces to the following $6\tm 6$ matrix

\begin{equation}
\begin{array}{cccc}
 & {\begin{array}{ccccc}
\hs{0.8cm} 5_H\hspace{0.3cm} & \hspace{0.5cm}5_{\Si } \hspace{0.3cm}
& \hspace{0.5cm}45_{\Si }\hspace{0.3cm} &
\hspace{0.5cm}{\bf F}_{\De }\hspace{0.3cm}   &~
\end{array}}\\ \vspace{1mm}
\begin{array}{c}
\bar 5_H\\ ~\\ \vs{0.1cm} \bar 5_{\Si } \\ ~ \\ \ov{45}_{\Si }  \\ ~\\ {\bf \bar F}_{\De }
 \end{array}\!\!\!\!\! \hs{-0.1cm}&{\left(\begin{array}{cccccc}

\hs{0.3cm}0  &\hs{1.1cm}0  &\hs{1cm}0&
\\
 &&&
 \\
\hs{0.3cm}0  &\hs{1cm}0  &\hs{1cm}0&
 \hs{1cm}{\bf \Om }
\\
 &&&
\\
\hs{0.3cm}0  &\hs{1.1cm}0  &\hs{1cm}0&
\\
 &&&
\\
 ~ & \hs{1.1cm}{\bf \bar{\Om }} &  ~& \hs{1cm}{\bf M_F}
\end{array}\hs{-0.1cm}\right)}~,
\end{array}  \!\!
\label{M-DT1}
\end{equation}
which reproduce the results already discussed briefly in the previous section.

From Eq. (\ref{M-DT1}) we see that the triplets gain masses
from the integration of ${\bf F_{\De }}$ states with the entries ${\bf \Om }$, ${\bf \bar{\Om }}$ being crucial. Thus, the $3\tm 3$  induced mass matrix for the triplets
will have form
\beq
M_T\propto {\bf \Om_T}({\bf M_{F, T}})^{-1}{\bf \bar{\Om }_T}~,~~~~{\rm with}~~~~{\bf \Om_T}\propto {\bf \Om }~,~~{\bf \bar{\Om }_T}\propto {\bf \bar{\Om }}~,
\la{M-Tripl}
\eeq
where the subscript $T$ indicates that the appropriate matrices should be derived from matrices appearing in Eq. (\ref{M-DT1}).
For example ${\bf M_{F, T}}\simeq {\bf M_{F}}$ and for ${\bf \Om_T}, {\bf \bar{\Om }_T}$ one should  take into account some GUT Clebsch factors.
These factors do not play any  role in our analysis. It is important that the triplet $3\tm 3$ mass matrix is generated and all the triplets acquire
masses. The situation differs for the doublet fragments. Since the $50$-plets do not include the doublets, the appropriate ${\bf M_{F, D}}$,
${\bf \Om_D}$ and ${\bf \bar{\Om }_D}$ matrices will have dimensions $2\tm 2$, $3\tm 2$ and $2\tm 3$ respectively. Up to some irrelevant Clebsch factors its structure is
\beq
\begin{array}{ccc}
 & {\begin{array}{ccc}
 & &
\end{array}}\\ \vspace{1mm}
{\bf \Om_D }\propto
 &{\left(\begin{array}{cc}

\hs{0.3mm} \Phi_1+\Phi_{24}&\hs{0.1cm}\Phi_{24}+\Phi_{75}
\\
\vs{-0.4cm}
\\
 \hs{0.3mm}\Phi_1+\Phi_{24} & \hs{0.1cm} \Phi_{24}+\Phi_{75}
 \\
 \vs{-0.4cm}
 \\
 \hs{0.3mm} \Phi_{24}+\Phi_{75}&\hs{0.1cm}\Phi_1+\Phi_{24}+\Phi_{75}

\end{array}\right)},~~~
{\bf \bar{\Om }_D }\propto {\bf \Om_D }^T~,~~~{\bf M_{F, D}}\simeq \lan X\ran {\rm Diag}\l 1~,1\r
\end{array}
\la{Om-MF-Doubl}
\eeq
and the induced $3\tm 3$ doublet mass matrix is
\beq
M_D\propto {\bf \Om_D}({\bf M_{F, D}})^{-1}{\bf \bar{\Om }_D}~.
\la{M-Doubl}
\eeq
Clearly, due to the form of the matrices in Eq. (\ref{Om-MF-Doubl}), the matrix in Eq. (\ref{M-Doubl}) has one zero eigenvalue. The reason is simple:
this $3\tm 3$ mass matrix is effectively induced by integrating out two heavy states (the matrix ${\bf M_{F, D}}$ is of $2\tm 2$ dimension). Thus, one doublet pair is light,
and should be identified to the MSSM Higgs doublets. Once more we stress that this is a result in both {\bf (a)} and {\bf (b)} cases.

Let us summarize the role of the anomalous ${\cal U}(1)$ symmetry used here. It forbids the renormalizable coupling
$H^2, \Si^2, \Phi H\Si $ which would contribute to the MSSM doublet mass. This ${\cal U}(1)$ symmetry also guarantees  that $\lan \De \ran =0$. 
This is
important because the combination $\De \ov{\De }$ has negative ${\cal U}(1)$ charge and the allowed operators such as $\ov{\De }\De \l H^2+\Si^2\r$,
which might be induced by Planck scale physics, 
do not give rise to any contribution to the doublet mass. The condition $\lan \De \ran =0$ also guarantees that there are no mixings between $5_{\Phi }$ and
$5_{H, \Si }, 45_{\Si }$ states (see Eq. (\ref{M-DT})) and thus the integration of heavy $5_{\Phi }, \ov{5}_{\Phi }$-plets does not destroy the DT hierarchy.

In what follows, we discuss some details of the Yukawa sector of this model.

\subsection*{Yukawa Sector and Fermion Mass Generation}

\begin{figure}[t]
\begin{center}
\hs{-0.3cm}
\resizebox{0.8\textwidth}{!}{
  \vs{-3cm}\includegraphics{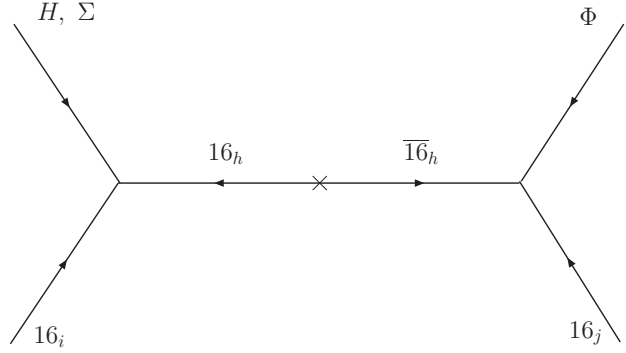}
}
\vs{-13cm}
\caption{Diagram inducing operators of Eq. (\ref{Phi-16-16-HSi}).}
\label{fig:1}       
\end{center}
\end{figure}

Now we discuss the fermion sector of the model and show that the charge assignments given in
Table \ref{t:1} give a self--consistent picture. For the matter $16_i$-plets ($i=1,2,3$) we take
the family universal ${\cal U}(1)$ charge $Q_{16_i}=-1/2$. Then the Yukawa couplings are
\beq
\sum_{k=0}\l \fr{\Phi }{M}\r^k16_i16_jH+\sum_{k=0}\l \fr{\Phi }{M}\r^k16_i16_j\Si ~,
\la{16-Yuk}
\eeq
where $M$ is some cut-off scale and can be taken close to $M_{\rm Pl}$.
Note that without using $\Phi $ insertion, although both $H(10)$ and $\Si (120)$-plets include light Higgs doublets, only renormalizable couplings
$16\cdot 16H$ and $16\cdot 16\Si $ do not give desirable fermion mass pattern \cite{Lavoura:2006dv}. Thus, at least the first power of $\Phi $ in one of the couplings of Eq. (\ref{16-Yuk}) is needed. Note that the operators $\Phi 16\cdot 16H$, $\Phi 16\cdot 16\Si $ can be generated from renormalizable
couplings through integrating some heavy states. For example, introducing the heavy states $16_h$ and $\ov{16}_h$ with ${\cal U}(1)$ charges $-1/2$ and $1/2$,
the relevant couplings are
\beq
(H+\Si )16\cdot 16_h+\Phi 16\cdot \ov{16}_h+M_h16_h\cdot \ov{16}_h~.
\la{16-h-coupl}
\eeq
 Integration of $16_h, \ov{16}_h$ states induces the effective operators
 \beq
 \fr{\Phi }{M_h}16\cdot 16H+\fr{\Phi }{M_h}16\cdot 16\Si ~.
 \la{Phi-16-16-HSi}
 \eeq
This is shown in Fig. \ref{fig:1}.

\begin{figure}[t]
\begin{center}
\hs{-4cm}
\resizebox{0.8\textwidth}{!}{
  \includegraphics{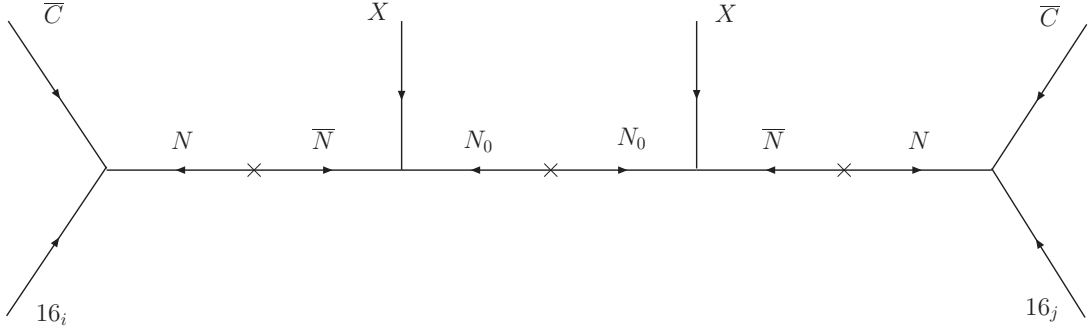}
}
\vs{-13.4cm}
\caption{Diagram generating operator of Eq. (\ref{Maj-a}).}
\label{fig:2}       
\end{center}
\end{figure}

Besides the Yukawa coupling discussed above we need the operator which will generate Majorana masses for the right handed neutrinos. For the case {\bf (a)}, the corresponding coupling is
\beq
\fr{X^2}{M_*^3}16_i16_j\bar C\bar C~.
\la{Maj-a}
\eeq
We now discuss the possibility of generating such couplings from renormalizable interactions. Introducing
$SO(10)$ singlet states $N, \bar N$ and $N_0$ with ${\cal U}(1)$ charges $2, -2$ and $0$ respectively, the allowed renormalizable
couplings are
\beq
N16\bar C+X\bar NN_0+M_NN\bar N+M_0N_0N_0~.
\la{ren-Maj-a}
\eeq
It is easy to check out that after integrating out the states $N, \bar N, N_0$, the operator in Eq. (\ref{Maj-a}) is generated with $M_*\sim (M_N^2M_0)^{1/3}$.  This integrating--out mechanism is shown in Fig. \ref{fig:2}.

\begin{figure}[t]
\begin{center}
\hs{-5cm}
\resizebox{0.85\textwidth}{!}{
  \includegraphics{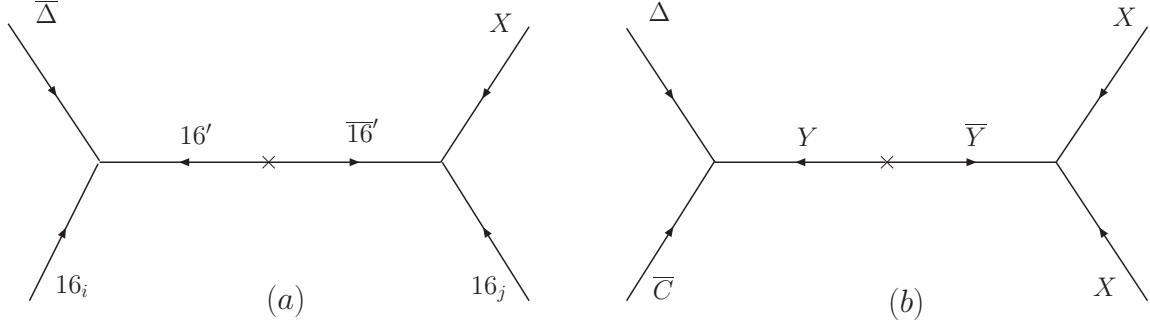}
}
\vs{-14cm}
\caption{Diagrams (a) and (b) generating operators of Eqs. (\ref{Maj-b}) and (\ref{W-De-C}) (case {\bf (b)}) respectively.}
\label{fig:3}       
\end{center}
\end{figure}

For case {\bf (b)} the coupling for the Majorana neutrino mass can be from the operator
\beq
\fr{X}{M_*}16_i16_j\bar{\De }~.
\la{Maj-b}
\eeq
Recall that $\bar{\De }$ can have the VEV [see Eq. (\ref{Sol-De-VEVs})] due the non-renormalizable coupling in Eq. (\ref{W-De-C}).
The operator in Eq. (\ref{Maj-b}) can also be generated from the renormalizable couplings. Introducing three pairs of $16'$, $\ov{16}'$-plets with ${\cal U}(1)$ charges $3/2$ and $-3/2$, the relevant superpotential terms are
\beq
16\bar{\De }16'+X16\ov{16}'+M_*16'\ov{16}'~.
\la{ren-Maj-b}
\eeq
Integration of $16', \ov{16}'$ states leads to the operator in Eq. (\ref{Maj-b}), with corresponding diagram in Fig. \ref{fig:3}.a. Besides this, we have to make sure that $\bar{\De }$ has a non--zero VEV.
For this to happen, the presence of the operator in Eq. (\ref{W-De-C}) [case {\bf (b)}] is important. If we wish
to not rely  on unknown Planck physics, these coupling can be generated by introducing
the scalar superfields $Y$ and $\bar Y$ with ${\cal U}(1)$ charges $4$ and $-4$ resp. With couplings $\bar C\De Y+X^2\bar Y+M_{\rm Pl}Y\bar Y$, the
integration of $Y, \bar Y$ states induce the operator {\bf (b)} in Eq.  (\ref{W-De-C}).
This is depicted in Fig. \ref{fig:3}.b.  Note that the additional field $\bar Y$ with negative ${\cal U}(1)$
charge has no VEV and therefore is harmless for DT hierarchy.

As wee see, the presented missing doublet $SO(10)$ model(s) is fully consistent with realistic fermion masses
and mixings. The remarkable thing is that the whole
scenario including the fermion sector can be constructed from renormalizable couplings.

\section{Discussion}

In this paper we have proposed a new solution of the doublet-triplet splitting problem
within $SO(10)$ GUT via a missing partner mechanism. For this mechanism to be realized through renormalizable superpotential couplings we have considered the scalar superfield content $10+120+126+\ov{126}+210$ and the $SO(10)$ rank breaking states $C, \bar C$. For the latter, two possibilities
{\bf (a)}: $C=16, \bar C=\ov{16}$ and {\bf (b)}: $C=126, \bar C=\ov{126}$ can be considered with equal success. Our scenario is consistent with realistic fermion sector as well as with successful gauge coupling unification.
Unification is achieved because below the GUT scale, the light fields are just those of the MSSM. One can also address the issue of gauge coupling perturbativity above the GUT scale. In this respect, let us point out that the chance is not bad. For instance, in case {\bf (a)}, one can consider the $SO(10)$ breaking down to $SU(5)$ at scale
$M_{SO(10)}\simeq \lan C\ran \simeq  \lan \bar C\ran\sim 10^{17}$~GeV. Below this scale, light scalar states which are needed are fragments from $10, 120$ and $75$ (from $210$). Note that the states $126, \ov{126}$ can have mass$\sim 10^{17}$~GeV and similar masses for the remaining fragments from $210$
(apart from the $75$-plet). All these can be achieved by a (mild) fine--tuning. Eventually, the VEV of $24$-plet (from $210$) is somewhat suppressed, but this does not change anything for the considered DT splitting scenario. With this mass spectrum (including light fermion families), the $SU(5)$ gauge coupling interpolated from $M_{SU(5)}\simeq 2\cdot 10^{16}$~GeV up to the  $M_{SO(10)}\sim 10^{17}$~GeV is still perturbative
$\al_{\rm GUT}(M_{SO(10)})\simeq 1/12.5$. Above the scale $M_{SO(10)}$, all $SO(10)$ states listed above should be included in the RGE study and one finds that the gauge coupling becomes strong near $1.7\cdot 10^{17}$~GeV. Thus, this scale should be considered as a natural cut--off of the theory.
Of course, more detailed study with accurate calculation of the mass spectrum is needed. Besides this question one  should also address proton stability and the problem of fermion flavor (mass and mixing hierarchy) within this scenario.

Since the mechanism which we have proposed opens up a wide playground for $SO(10)$ model building, we hope that our proposal will motivate others to address and investigate an array of issues which  we have not
attempted in this work.

\vs{0.5cm}

\hs{-0.6cm}{\bf Acknowledgments}

\vs{0.2cm}
\hs{-0.6cm}Discussions with S. Barr, Z. Chacko  and Q. Shafi  are acknowledged. The work is supported in part by DOE grant DE-FG03-98ER-41076 (K.S.B. and Z.T.) and DE-FG02-91ER40626  (I.G.).

\bibliographystyle{unsrt}

\newpage

\end{document}